\documentclass{elsart}
\journal{New Astronomy}

\usepackage{natbib}

\def\kev{~{\rm keV}}

\def\beq{\begin{equation}}
\def\eeq{\end{equation}}
\def\bea{\begin{eqnarray}}
\def\eea{\end{eqnarray}}
\usepackage{aas_macros}
\usepackage{graphicx}
\usepackage{epsfig}

\usepackage{amssymb}

\begin{document}

\begin{frontmatter}



 \title{Intracluster Medium through three years of WMAP}


\author{Niayesh Afshordi}
\address{Institute for Theory and Computation, Harvard-Smithsonian Center for Astrophysics, MS-51, 60 Garden Street, Cambridge, MA 02138, USA\\
E-mail: nafshordi@cfa.harvard.edu}

\begin{abstract}
Wilkinson Microwave Anisotropy Probe (WMAP) has provided us with the
highest resolution all-sky maps of the Cosmic Microwave Background.
As a result of thermal Sunyaev-Zel'dovich effect, clusters of
galaxies are imprinted as tiny, poorly resolved dips on top of
primary CMB anisotropies in these maps. Here, I describe different
efforts to extract the physics of Intracluster Medium (ICM) from the
sea of primary CMB, through combining WMAP with low-redshift galaxy
or X-ray cluster surveys. This finally culminates at a mean
(universal) ICM pressure profile, which is for the first time
directly constrained from WMAP 3yr maps, and leads to interesting
constraints on the ICM baryonic budget.

\end{abstract}

\begin{keyword}
cosmology \sep cosmic microwave background \sep methods: data
analysis \sep intergalactic medium


\end{keyword}

\end{frontmatter}

\section{Introduction}
Since the discovery of dark matter by Zwicky in 1937
\cite{1937ApJ....86..217Z}, which was done through study of virial
velocities of galaxies in clusters, galaxy clusters have been in the
forefront of cosmological studies.

One of the most promising tracers of galaxy clusters is the thermal
Sunyaev-Zel'dovich (SZ) effect, induced in the Cosmic Microwave
Background (CMB) sky through scattering of CMB photons off hot
electrons in the Intracluster Medium (ICM) \cite{sunyaev72}. Most
significant properties of SZ effect as a tracer of clusters can be
summarized as \cite{2002ARA&A..40..643C}:
\begin{itemize}
\item It is proportional to the line of sight integral of the
ICM pressure.
\item It has a characteristic frequency dependent signature which
enables observers to extract it from the CMB background.
\item The total SZ flux of a cluster does not depend on redshift for
$z\gtrsim 1$, making SZ effect an ideal tool for making high
redshift cluster surveys.
\end{itemize}

In contrast to free-free X-ray emission, which traces the square of
electron density, the fact that SZ effect follows gas pressure makes
it a more representative tracer of gas density, and less sensitive
to small scale structure or astrophysical processes (e.g.
\cite{2006astro.ph..1133R}). Therefore, it is believed that SZ flux
can be used as a reliable tracer of cluster mass with relatively low
scatter.

This has motivated many SZ cluster surveys, which are expected to
give a census of dark matter halo mass function at high redshifts,
and thus enable novel constraints on the expansion history of the
Universe \cite{2006astro.ph..5575B}.

However, despite theoretical expectations, it is not quite clear how
well SZ observations can be calibrated to trace dark matter halo
properties. Therefore, complementary X-ray and SZ observations, as
well numerical studies are necessary to obtain a realistic picture
of the ICM properties.

In this note, I outline efforts to extract the SZ effect from the
highest resolution all-sky maps of CMB produced by Wilkinson
Microwave Anisotropy Probe (WMAP) \cite{bennett03,hinshaw06}, the
resulting constraints on ICM pressure profile, and comparison with
other X-ray and SZ observations, as well as numerical studies.

\section{Optimal SZ detection in cross-correlation}

\begin{figure}[!t]
\centerline{\includegraphics[width=0.8\linewidth]{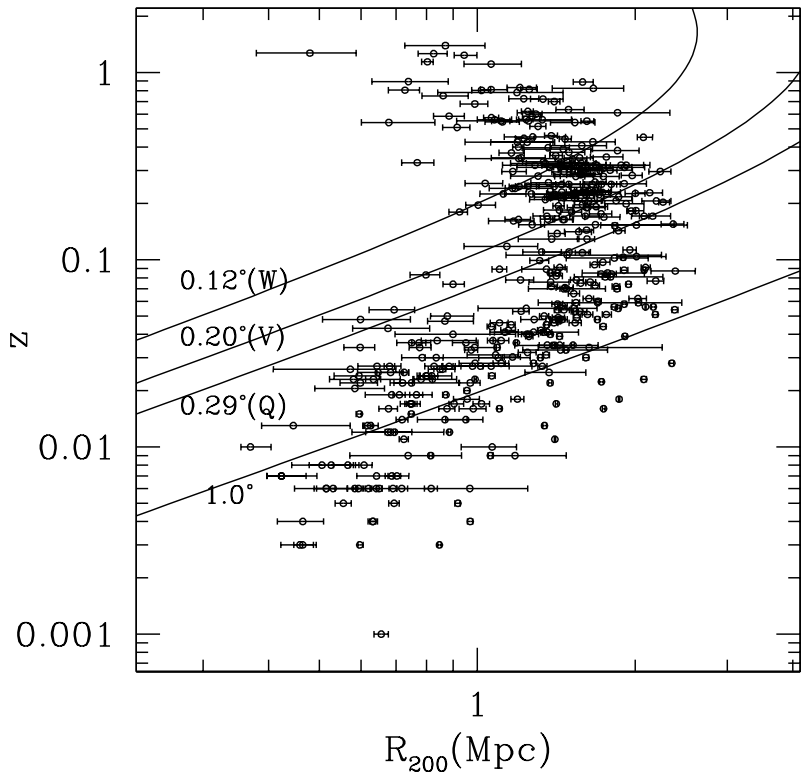}}
\caption{Estimated virial radii, $R_{200}$, for all clusters with
measured X-ray temperatures in the literature \cite{ALNS}, plotted
against their redshift. The lines show the physical sizes of the
effective angular radii of WMAP beams, as well as  a degree angular
scale, as a function of redshift in a $\Lambda$CDM cosmology. Many
low-redshift clusters are in fact resolved by WMAP \cite{ALS}.}
\label{zrvir}
\end{figure}

While SZ observations of a limited number of massive clusters are
existent in the literature
\cite{2006astro.ph..4039L,2004ApJ...617..829B}, due to their limited
resolution and sensitivity, they have only been used to put
constraints on over-all SZ flux (or gas mass) in conjunction with
X-ray observations. Although WMAP has a worse resolution than the
instruments commonly used to carry out SZ observations, it covers
the whole sky, and thus includes SZ signatures of all the galaxy
clusters in the Universe (see Fig. \ref{zrvir}). Therefore, although
the WMAP SZ signals of individual clusters have a low significance,
it is possible to combine SZ signatures of many clusters to obtain
constraints on the mean ICM properties.

\begin{figure}[!t]
\centerline{\includegraphics[width=0.6\linewidth]{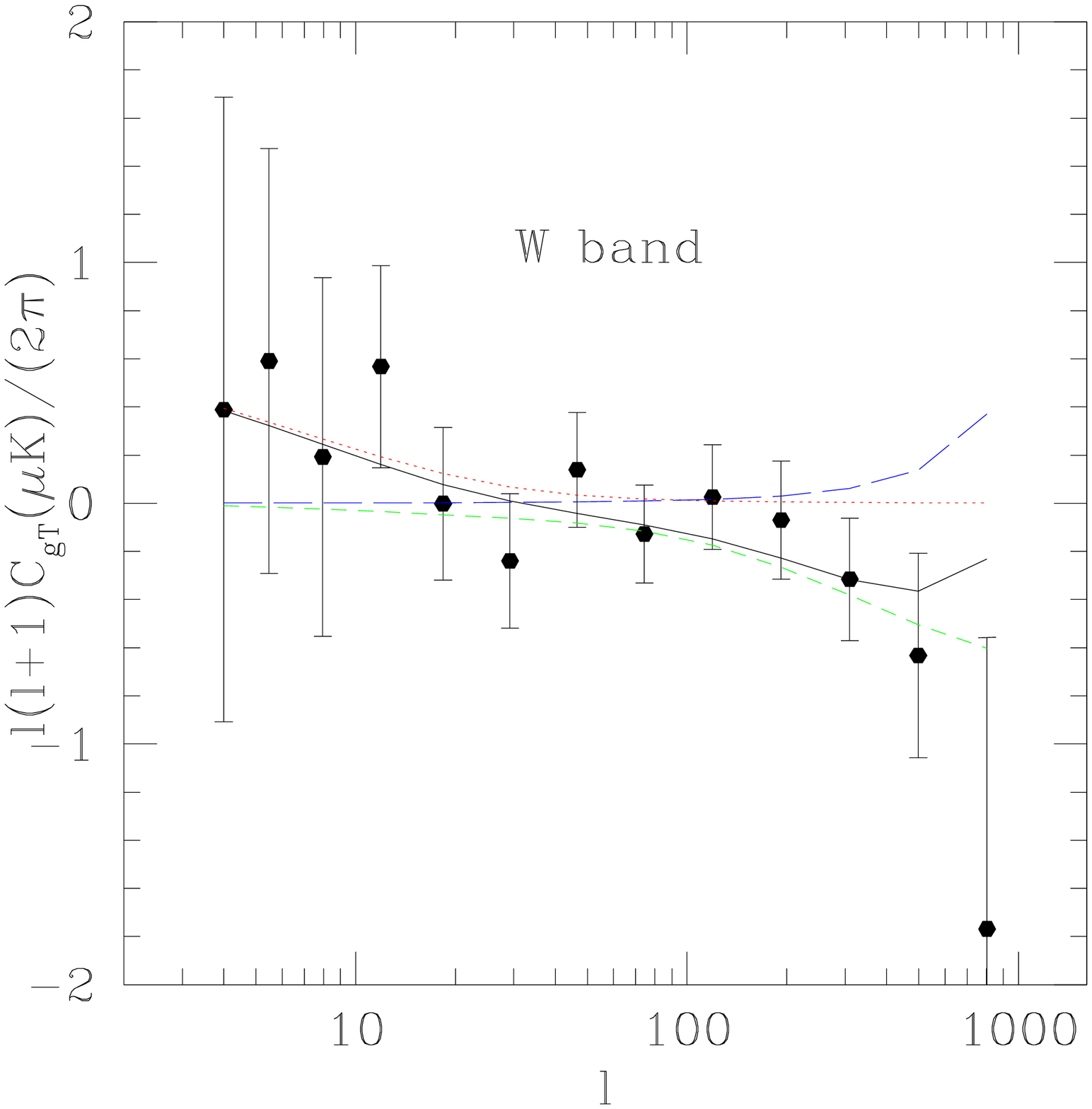}}
\caption{The cross-power spectrum of 2MASS XSC galaxies in the
magnitude bin $13.5 < K_{20} < 14$, with the 1st yr WMAP W-band
\cite{2004PhRvD..69h3524A}. The red (dotted), blue (long-dashed),
and green (short-dashed) lines correspond to best fit Integrated
Sachs-Wolfe, radio point source, and thermal SZ contributions
respectively. The marginalized significance of the thermal SZ signal
(4 2MASS magnitude bins x 3 WMAP frequency bands) is $3.7\sigma$.}
\label{2MASS}
\end{figure}

First efforts in this direction have been made through direct
cross-correlation of galaxy/cluster surveys with the WMAP
temperature maps
\cite{bennett03,2003ApJ...597L..89F,2004MNRAS.350L..37F,2004MNRAS.347L..67M,2004PhRvD..69h3524A}.
In particular, \cite{2004PhRvD..69h3524A} finds an almost $4\sigma$
detection of the thermal SZ effect through analysis of 2MASSxWMAP
cross-power spectrum (Fig. \ref{2MASS}). While these results were
generally consistent with theoretical expectations, the observable
signal contains a mixture of information, including non-linear
galaxy bias, halo clustering, and ICM physics. Moreover, the
non-Gaussian/non-linear nature of the SZ signal implies that a
simple cross-power spectrum analysis is not the optimum way to
unleash its full statistical significance. This became clear
following the analysis of
\cite{2004MNRAS.347..403H,2004ApJ...613L..89H} who could find $S/N
\sim 5$ for their SZ detection, through cross-correlating with a
non-linear template made out of different galaxy surveys. Despite
its better statistical significance, the disadvantage of the latter
method was that the galaxy biasing was mixed with ICM physics in a
yet more complicated way, making comparison with theory even less
certain (but see \cite{2006astro.ph..6172H}).

Is there an optimum way to extract the SZ signal of a low-resolution
CMB map, using a survey of the low redshift universe, without mixing
ICM physics and non-linear clustering information? Although there
have been many studies of the correlation of optical properties of
clusters with their dark matter halo properties, such correlations
have been far from perfect. On the other hand, the temperature of
the ICM diffuse X-ray emission has been shown to be a far better
tracer of halo intrinsic properties\footnote{One may worry about the
cyclic nature of this statement, as different physical properties
(such as virial mass, gas mass, and X-ray temperature) are derived
from the same X-ray observations.}(e.g. see
\cite{2002ApJ...564..669A}, and references therein).

\begin{figure}[!t]
\centerline{\includegraphics[width=0.8\linewidth]{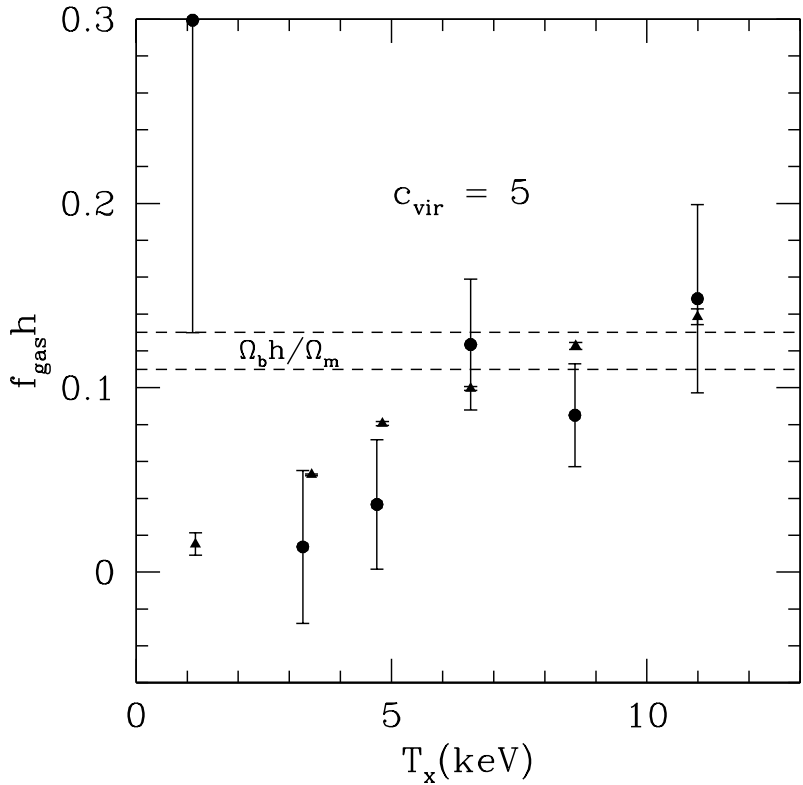}}
\caption{The inferred ICM gas mass fraction times $h$ (Hubble
constant in units of $100~ {\rm km/s/Mpc}$), averaged within $2~
{\rm keV}$ $T_{\rm x}$-bins \cite{ALS}. The solid dots correspond to
WMAP 1st yr SZ signal, while triangles are gas fractions inferred
from cluster X-ray luminosities. The dark matter halo is assumed to
have an NFW profile \cite{nfw} with $c_{\rm vir}=5$, and the dotted
lines show the cosmic baryonic budget.} \label{szmap1}
\end{figure}

This motivated our study of WMAP 1st year data \cite{ALS}, where we
constrained a mean (universal) model for ICM, assuming that all the
ICM properties have a minimal scaling with the observed X-ray
temperature of galaxy clusters. The universal ICM model in
\cite{ALS} assumes a polytropic ICM which is in hydrostatic
equilibrium within a dark matter NFW potential \cite{nfw}, and is
truncated by an accretion shock at the virial radius. This only
leaves the total ICM gas mass fraction (or $f_{\rm gas}$) as the
normalization of the expected SZ signal, which we can fit for as a
function of $T_{\rm X}$, by looking at WMAP maps within the
neighborhood of a compilation 116 X-ray clusters (Fig.
\ref{szmap1}). After marginalizing over possible point source
contamination, this leads to an $8\sigma$ detection of SZ effect.

\section{Reconstructing the ICM mean pressure profile}

Due to their limited dynamical range, none of the SZ observations of
galaxy clusters have so-far been able to significantly constrain the
ICM pressure profile. As a result, the interpretation of SZ
observations have always relied on profiles that are either directly
observed in X-rays (e.g.
\cite{2004ApJ...617..829B,2006astro.ph..4039L}), or are motivated by
X-ray observations (e.g. our analysis of \cite{ALS}, discussed
above). As pointed out by
\cite{2004ApJ...617..829B,2004MNRAS.352.1413S}, this could
potentially lead to systematic discrepancies, especially if X-ray
and SZ observations probe different parts of the ICM profile (e.g.
\cite{2005astro.ph.10160L}). Moreover, as X-ray emission is a
non-linear function of gas density, any inference of a gas
density/pressure profile is model-dependent and is thus prone to
systematic biases.

Given the large dynamical range of WMAP resolution, and the high
$S/N$ achieved in our earlier analysis \cite{ALS}, it is conceivable
that one could construct an ICM pressure profile by effectively
stacking all the clusters that are resolved by WMAP. Following this
line of thought, in \cite{ALNS}, we devise a method to constrain the
full mean (or universal) ICM pressure profile using the newly
released WMAP 3yr CMB maps. The only remaining theoretical
assumption in this approach is that the physical size of the ICM
profile scales as $T_{\rm X}^{1/2}$, which is consistent with
numerical simulations \cite{2006astro.ph..3205K} and high resolution
X-ray observations \cite{2006ApJ...640..691V}.

\begin{figure}[!t]
\centerline{\includegraphics[width=0.8\linewidth]{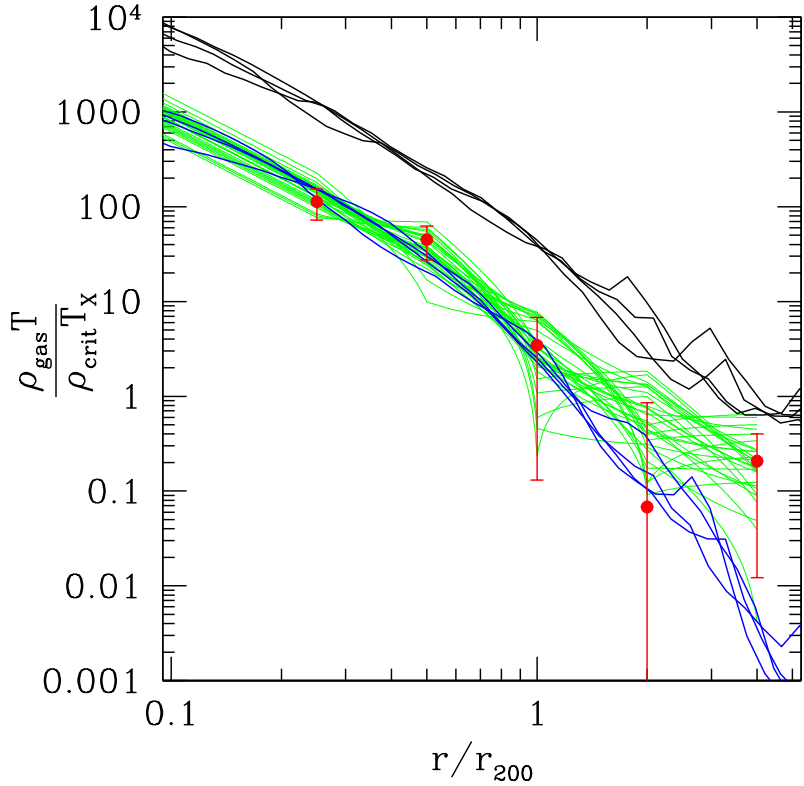}}
\caption{The mean pressure profile of 109 of our most massive
clusters with $T_X > 5\kev$ (red points+errorbars); \cite{ALNS}. The
green curves are 30 random realizations of the measured points,
which reflect the errors, as well as their correlations. The blue
curves are predicted pressure profiles from four simulated clusters
with $T_X > 5\kev$ \cite{kravtsov_etal05, nagai_etal06}, while black
curves show dark matter density from the same simulations, divided
by the critical density of the Universe, $\rho_{\rm crit}$.}
\label{prof}
\end{figure}

Fig. (\ref{prof}) shows the mean pressure profile for our most
massive clusters, with $T_X > 5~{\rm keV}$, after we exclude
clusters with significant point source contamination. We see that
the reconstructed pressure profile (red points/green curves) are in
excellent agreement with realistic hydrodynamic simulations
\cite{kravtsov_etal05, nagai_etal06}, in the same temperature range.

\section{Are we missing cluster baryons?}

One of the most surprising findings of recent SZ observations (e.g.
\cite{2006astro.ph..4039L}, and references therein) which is
confirmed by our WMAP SZ observations at larger scales \cite{ALS,
ALNS} is that about $30-40\%$ of cluster baryonic budget is missing
from the Intracluster Medium. In fact, WMAP maps are the first to
have enough dynamical range to show that these baryons cannot be
present within the virial radius of ICM. Similar results are also
seen in recent X-ray observations \cite{2003MNRAS.344L..13E,
2006ApJ...640..691V}, while hydrodynamic simulations contain a
similar gas fraction in the form of cold gas \cite{nagai_etal06}.

On the other hand, stellar mass of cluster galaxies only seems to
account for $10-15\%$ of cluster baryonic budget
\cite{2003ApJ...591..749L}, leaving about $15-30\%$ of the cluster
baryonic budget (at $2-3\sigma$ level) un-accounted for. There does
not seem to be a conventional solution to this new dilemma of the
ICM physics \cite{ALNS}.

\section{Conclusions and Outlook}

As the era of the next generation of CMB experiments is fast
approaching, it is vital to understand the kind of signals that
these experiments are expected to see, the most significant of which
is the thermal SZ effect. Moreover, the SZ effect is believed to be
the most efficient and robust way of detecting clusters at high
redshifts, which can be used to constrain the history of dark energy
in that era. To what extent these promises can be realized is the
subject of on-going research and inquiry.

In this note, we argued that despite its low resolution, WMAP, as
the highest resolution existing all-sky CMB map, can bring in a
unique perspective into the understanding of cluster SZ properties,
and ICM physics in general. However, this information can only be
extracted through combination with a tracer of low redshift
universe, such as galaxy surveys or X-ray cluster observations. We
described different methods to exploit this SZ signal, which involve
varying degrees of theoretical modeling. In particular, WMAP
provides the only available observations of the ICM outskirts, and
thus for the first time constrain the cluster total SZ flux (or
thermal energy).

 Both X-ray and SZ observations of the
ICM are prone to systematics and technological limitations that
could complicate or confuse their interpretation. While on-going
theoretical/numerical studies go a long way in understanding these
systematics, it would not be the first time that nature outdid the
Astronomers' imagination. Therefore, complementary cluster
observations (and surveys) in SZ, X-rays, and gravitational lensing
(along with numerical studies) is the only guaranteed way to obtain
a comprehensive picture of the ICM physics, and thus be able to
reliably use them as a {\it Dark Energy machine}.




The topics discussed here have been borne out of past, or on-going
collaborations with Yen-Ting Lin, Yeong-Shang Loh, Daisuke Nagai,
Alastair Sanderson, and Michael Strauss. I would like to thank them
for continuous scientific discourse and education.

\bibliography{szmap_irvine}
\end{document}